\documentclass[aps,preprint]{revtex4-1}

\usepackage{amsmath,amsfonts,amssymb,mathbbol,dsfont}
\usepackage{graphicx,psfrag,color}
\usepackage{dcolumn}
\usepackage{bm}
\usepackage{mathbbol, appendix}

\def\r{{\bm{r}}} 
\def\x{{\bm{x}}} 
\def\y{{\bm{y}}} 

\def\one{{\mathbb{1}}}
\def\im{{\sf {i}}}

\begin{document}

\title[]{Modeling Electron Fractionalization 
with Unconventional Fock Spaces}

\author{Emilio Cobanera
\footnote{Present address: Department of Physics and Astronomy, Dartmouth
College, 6127 Wilder Laboratory, Hanover, NH 03755, USA} 
}

\affiliation{Institute for Theoretical Physics, Center for Extreme Matter 
and Emergent Phenomena, Utrecht University, Leuvenlaan 4, 3584 CE Utrecht, The Netherlands}


\begin{abstract}
It is shown that certain fractionally-charged quasiparticles can be 
modeled on \(D-\)dimensional lattices in terms of unconventional yet 
simple Fock algebras of creation and annihilation operators. These 
unconventional Fock algebras are derived from the usual fermionic algebra 
by taking roots (the square root, cubic root, etc.) of the usual 
fermionic creation and annihilation operators. If the fermions carry 
non-Abelian charges, then this approach fractionalizes the Abelian charges 
only. In particular, the \(m\)th-root of a spinful fermion carries charge \(e/m\) 
and spin \(1/2\). Just like taking a root of a complex number, taking a 
root of a fermion yields a mildly non-unique result. As a consequence,
there are several possible choices of quantum exchange statistics for 
fermion-root quasiparticles. These choices are tied to the 
dimensionality \(D=1,2,3,\dots\) of the lattice by basic physical considerations. 
One particular family of fermion-root 
quasiparticles is directly connected to the parafermion zero-energy modes 
expected to emerge in certain mesoscopic devices involving fractional quantum 
Hall states. Hence, as an application of potential mesoscopic interest, I 
investigate numerically the hybridization of Majorana and parafermion zero-energy 
edge modes caused by fractionalizing but charge-conserving tunneling.
\end{abstract}


\maketitle

\section{Introduction}

Lattice-regularized theories are everywhere in physics, because 
they capture in a transparent way the interplay between internal 
symmetries, strong correlations \cite{wilson1974}, topological 
quantum orders \cite{wen2002,kitaev2003,nussinov2009}, dualities
\cite{savit1980,cobanera2011}, and disorder \cite{anderson,ryu2010}.
They represent a rich class of models naturally suited for numerical 
simulations, notwithstanding the fact that they pose exponentially 
hard computational problems more often than not. Quantum 
information processing tools are best adapted to lattice-regularized 
systems \cite{amico2008}. Because microscopically motivated lattice 
theories can be extraordinarily detailed, it is often necessary or 
desirable to search for simplifications. The schematic models
that result, sometimes derived and sometimes conjectured, are regarded 
as effective, and one seeks to incorporate in them as much as one 
believes to know about the nature of the quasiparticle excitations 
dominant in the energy scales of interest.
  
This paper focuses on one particular topic where effective lattice 
models have played a relatively small role thus far: electron 
fractionalization. Consider for concreteness the Su-Schrieffer-Heeger 
(SSH) model \cite{su1979} of electrons coupled to phonons in one 
dimension. In this model of the linear polyacetylene molecule, kinks 
in the phonon field carry a fractional amount of electric charge, due 
to dressing by the electronic degrees of freedom. Typically, this 
dressing mechanism is investigated in an approximation in which 
the kinks are static. This approximation is very fruitful and 
admits grand generalizations in field theory \cite{niemi}. 
But in a less simplified picture where the phonons are not frozen, 
the electrons might coexist with dynamic kink excitations, and channels 
for electron breakup into fractionally charged kinks and recombination 
might perhaps exist. If this were the case, an {\it effective} 
lattice field theory of such a regime, especially a simple one, would 
be quite different from the starting point, the SSH model.
Presumably it would feature effective creation and annihilation 
operators of dressed kinks. But what should be the algebra of such 
effective creation and annihilation operators?

While the SSH model is arguably the simplest model of charge 
fractionalization, the idea of dynamic, fractionally charged
quasiparticles has been pursued much further in the context of 
fractional quantum Hall (FQH) effects \cite{dassarma1997,heinonen1998}. 
In these systems, the exact microscopic origin of fractionally-charged 
quasiparticles is still an active area of research \cite{murthy2003}. 
However, for the FQH states characterized by Jain fractions \(\nu=p/q\) 
(\(q\) must be an odd integer), there is agreement on the nature of 
these quasiparticles on three basic points. First, the charge of 
the elementary \cite{foot1} quasiparticles is \(e^*=e/q\), independently 
of \(p\) (\(e<0\) is the charge of the electron). Second, these 
quasiparticles display unconventional exchange quantum statistics 
characterized by the statistical angle 
\(\theta=e^{i\pi/q}\) \cite{su1986}. Third, an electron entering a FQH 
fluid may undergo charge breakup, splitting into several mobile, if it were
not for disorder, quasiparticles, and, conversely, these quasiparticles 
may recombine into an electron \cite{goldhaber2003}. The second point,
unconventional quantum statistics, is hard to probe experimentally. Shot 
noise experiments are good evidence for the first and third 
points \cite{shot}. One can summarize these three basic ideas by saying 
that the electron is fractionalized in the FQH liquids associated to the 
Jain fractions, in the very precise sense that it breaks up into 
indistinguishable quasiparticles that behave as fractions of the
original electron.  

In this paper it will be shows that it is possible to capture the
three basic physical features of a ``fraction of an electron" in terms 
of unconventional Fock algebras of creation and annihilation operators.
These unconventional Fock operators create/annihilate a fraction of a 
fermion in a surprisingly literal sense, and can be used for describing 
a fraction of a spinful fermion and, more generally, a fraction of a 
fermion carrying any sort of non-Abelian internal symmetry. The unconventional 
Fock algebras of this paper can be chosen to display anyonic features, 
but are not mathematically tied to space dimensionality, they can even 
adjusted to be physically sound in more than two space dimensions. 
As a simple but suggestive application, I investigate numerically the 
unconventional Josephson effect in a model 
of a parafermion chain \cite{fendley2012} (roughly, a system designed to 
emerge on the edge of an FQH liquid \cite{lindner2012}) coupled to a Majorana 
chain \cite{kitaev2001,mourik2012} by charge-conserving tunneling of spinless 
electrons. In one dimension, and in the continuum, these systems can also be 
described with the aid of bosonization \cite{alicea2013}.
However, unlike the formalism of this paper, bosonization does not grant 
direct numerical access to the problem considered here.    

The direct link between the formalism of this paper and traditional 
approaches to the FQH effect has yet to be established. There is however a 
potentially significant indirect connection already in place. The fractionalized 
fermions of this paper directly linked to parafermions, and parafermions have 
in turn been recently linked to the FQH effect \cite{burrello2013,pfsbib,mong2014}. 
The relationship can be suspected from the fact that parafermions fractionalize 
Majorana modes \cite{lindner2012}, while in this paper the very fermions 
are fractionalized by comparable algebraic means. A second-quantized description 
of parafermions was introduced in Ref.\,\cite{cobanera2014}, but the relationship 
of that formalism to fermion fractionalization was not recognized at the time. Our 
main focus here is on describing a more general, direct approach to modeling by 
creation and annihilation operators a fraction of a fermion carrying a non-Abelian 
charge like spin, in a way consistent with a lattice regulator and not restricted 
to one dimension.

\section{Fermion-root quasiparticles}

Imagine a charge-conserving breakup/recombination process \(e \leftrightarrow  me^*\)
in which a fermion of charge \(e\) decays into \(m\) indistinguishable
quasiparticles of charge \(e^*=e/m\), or \(m\) quasiparticles recombine
into a fermion. The integer \(m=2,3,\dots\) is the number of quasiparticles 
that are produced when the fermion undergoes charge breakup because it left 
a normal medium and entered a fractionalizing one, for example. How should one 
effectively describe this process, in second quantization and on a lattice? 
Because of the lattice regularization, the process \(e \leftrightarrow  me^*\) 
may be modeled as taking place at one site, or more precisely, at one 
single-particle state \(\alpha\). Then \(e \leftrightarrow  me^*\) 
translates into the equations 
\begin{eqnarray}\label{ansatz}
C_\alpha^{\dagger m}=f_\alpha^\dagger, \quad C_\alpha^m=f_\alpha.
\end{eqnarray}
These equations state that creating or annihilating \(m\) quasiparticles
in the state \(\alpha\) is the same thing as creating or annihilating a 
fermion in that state. Thus, the task is to find quasiparticle operators 
\(C_\alpha, C_\alpha^\dagger\) such that a) their \(m\)th powers satisfy 
canonical anticommutation relations, and b) they have a second quantization 
interpretation of their own. The quasiparticles they create or annihilate
are the {\it fermion-root quasiparticles} of this paper. 

It is possible to split the task of solving Eq.\,\eqref{ansatz} into two 
steps. Let us examine first the case when there is only one single-particle 
state, that is, only one label \(\alpha\). Then one need not worry about 
exchange statistics. Only exclusion statistics \cite{batista} and the more 
technical problem of normal ordering matter. The exclusion statistics of the 
fermion-root quasiparticles follows from Eq.\,\eqref{ansatz}, since then 
\begin{eqnarray}\label{exc}
C_\alpha^{\dagger\,2m}=(C_\alpha^{\dagger\, m})^2=f_\alpha^{\dagger\, 2}=0,\quad
C_\alpha^{2m}=(C_\alpha^m)^2=f_\alpha^2=0.
\end{eqnarray}
This means that the state \(\alpha\) can be occupied by at most by \(2m-1\) 
quasiparticles. It is also necessary to be able to normal-order quasiparticle
creation and annihilation operators. The rule for exchanging creation and
annihilation operators is
\begin{eqnarray}\label{no}
C_\alpha^lC_\alpha^{\dagger l}+C_\alpha^{\dagger (2m-l)}C_\alpha^{2m-l}= \mathds{1},
\quad\quad l=1,\dots, 2m-1.
\end{eqnarray}
Then, as a consequence, the associated ``composite" fermion satisfies the 
canonical relation
\begin{eqnarray*}
C_\alpha^mC_\alpha^{\dagger m}+C_\alpha^{\dagger m}C_\alpha^m=
f_\alpha f_\alpha^\dagger+f_\alpha^\dagger f_\alpha=\one.
\end{eqnarray*}
 
There is a simple interpretation of Eq.\,\eqref{no}. The operator 
\(C_\alpha^{\dagger(2m-l)}C_\alpha^{2m-l}\) annihilates any 
state with less than \(2m-l\) particles, and \(C_\alpha^lC_\alpha^{\dagger l}\)
annihilates any state with \(2m-l\) or more particles. Then Eq.\,\eqref{no},
the rule for normal ordering quasiparticles and quasiholes, states that these 
two sets of states taken together conform the totality of the Fock space (since 
there is only one label \(\alpha\)). This interpretation also suggests that
\begin{eqnarray}\label{counting}
N_\alpha=\sum_{l=1}^{2m-1}C_\alpha^{\dagger l}C_\alpha^{l}
\end{eqnarray}
might play the role of number operator for the quasiparticles. To check that 
this is correct, one exploits the normal ordering and exclusion rules to show
that 
\begin{eqnarray}\label{u1}
[N_\alpha,C_\alpha^l]&=&-lC_\alpha^l,\qquad \quad l=1,\dots,2m-1.
\end{eqnarray}
In other words, \(N_\alpha\) generates \(U(1)\) rotations of the 
quasiparticle creation and annihilation operators. As a consequence, 
fermion-root quasiparticles can be counted and minimally coupled 
to gauge fields. Moreover, the Fock vacuum \(|0\rangle\) for the 
quasiparticles is defined by the usual condition \(C_\alpha|0\rangle=0\), 
since then \(N_\alpha|0\rangle=0\). The full quasiparticle Fock space
is spanned by the states \(|l\rangle=C_\alpha^{\dagger\,l}|0\rangle\)
with \(0\geq l<2m\) quasiparticles.  

Our discussion so far solves completely the problem of taking the
\(m\)th root of a single fermion. The solution displays
nice physical properties. If there is more than one single-particle
state, then one must in addition determine the exchange statistics of 
the fermion-root quasiparticles. At this point it becomes necessary to 
order the single particle labels, so let us index them with an integer \(i\). 
Let \(C_{\alpha_i}'\) denote a set of operators satisfying 
the exclusion and normal ordering rules, Eqs. \eqref{exc} and \eqref{no},  
{\it and commuting if they carry different labels}. The associated
number operators are \(N_{\alpha_i}'\). One can construct such 
a set by tensoring the \(2m\times 2m\) matrices
\begin{eqnarray*} 
{C}=
\begin{pmatrix}
0& 1& 0& \cdots& 0\\
0& 0& 1& \cdots& 0\\
0& 0& 0& \cdots& 0\\
\vdots& \vdots& \vdots&      & \vdots \\
0& 0& 0& \cdots& 1\\
0& 0& 0& \cdots& 0\\
\end{pmatrix}, 
\quad
{N}=
\begin{pmatrix}
0& 0& 0& \cdots& 0\\
0& 1& 0& \cdots& 0\\
0& 0& 2& \cdots& 0\\
\vdots& \vdots& \vdots&      & \vdots \\
0& 0& 0& \cdots& 0\\
0& 0& 0& \cdots& 2m-1\\
\end{pmatrix}
\end{eqnarray*}
with identity matrices. Hence, if there are \(N\) single-particle labels 
\(\alpha_i\), then the dimension of the underlying Hilbert space is 
\((2m)^{N}\). With these definitions, the quasiparticle creation and 
annihilation operators
\begin{eqnarray*}
C_{\alpha_i}=C_{\alpha_i}' e^{\im \theta \sum_{j<i} N_{\alpha_j}'},\quad
 C_{\alpha_i}^\dagger=(C_{\alpha_i}')^\dagger e^{-\im \theta \sum_{j<i} N_{\alpha_j}'},
\end{eqnarray*}
satisfy, thanks to Eq.\,\eqref{u1}, the exchange rules 
\begin{eqnarray}
\label{exch}
C_{\alpha_i}C_{\alpha_j}=e^{\im \theta }C_{\alpha_j}C_{\alpha_i},\quad
C_{\alpha_i}C_{\alpha_j}^\dagger =
e^{-\im \theta}C_{\alpha_j}^\dagger C_{\alpha_i},\quad i<j.
\end{eqnarray}

By construction, the statistical angle \(\theta\) for the exchange of 
two quasiholes (or quasiparticles) turns out to be the negative of that 
associated to the exchange of a quasihole with a quasiparticle. 
For the Laughlin series \(\nu=1/q\) of FQH liquids, the same result is 
obtained in a very different way from computing a Berry phase in the 
manifold of Laughlin's wave functions \cite{arovas1984,su1986}. 
Because in this paper quasiparticles are constructed so as to 
yield the \(m\)th root of a fermion, \(\theta\) must be such that   
\begin{eqnarray*}
C_{\alpha_i}^mC_{\alpha_j}^m+C_{\alpha_j}^m C_{\alpha_i}^m=
f_{\alpha_i}f_{\alpha_j}+f_{\alpha_j} f_{\alpha_i}=&\ 0,\\
C_{\alpha_i}^mC_{\alpha_j}^{\dagger\, m}+C_{\alpha_j}^{\dagger\, m} C_{\alpha_i}^m=
f_{\alpha_i}f_{\alpha_j}^\dagger+f_{\alpha_j}^\dagger f_{\alpha_i}=&\ 0,\quad\qquad i\neq j.
\end{eqnarray*}
According to Eq.\,\eqref{exch}, these conditions are equivalent to the 
single constraint \(e^{\im\theta m^2}=-1\) on the statistical angle. 
There are \(m^2\) solutions
\begin{eqnarray*}
\theta_\ell=\frac{\pi(2\ell+1)}{m^2}, \quad \ell=0,1,\dots,m^2-1.
\end{eqnarray*}
The various allowed values of \(\theta\) are qualified by physical 
considerations. At the most basic level, the {\it crucial difference}
between \(m\) even and \(m\) odd is that if \(m\) is odd, and only
in this case, then there exists one unconventional Fock algebra with exchange 
statistics \(e^{i\theta_\ell}=-1\), for \(\ell=(m^2-1)/2\). For lattices of 
dimension \(D\geq 3\), only \(\theta=\pi\) is allowed by locality \cite{foot3}, 
implying that only \(m\) odd is physically useful within the framework of this 
paper. From the point of view of Jain's composite fermion theory \cite{heinonen1998}
of the FQH liquids at odd denominators, it is reassuring that \(\theta=\pi\) 
is always allowed if \(m\) is odd. Composite fermions are, first and foremost, 
fermions whose charge has been renormalized non-adiabatically to a fractional 
value. It is this non-adiabatic renormalization of the charge that is unique 
to two dimensions \cite{goldhaber2003}. In any case, unlike the exchange angle, 
the exclusion and normal ordering rules are indifferent to the dimension of the 
lattice and whether \(m\) is even or odd.

Next I will focus on low (\(D=1,2\)) dimensional lattices, in order to allow
for various possible statistical angles. It is convenient to 
decompose \(m\) as \(m=2^r q\), with \(q\) odd. For simplicity, I will discuss 
first the case \(r=0\) and show how to adjust \(\theta\) in accordance to the 
quasiparticles of the FQH liquids at odd-denominator filling fractions 
\(\nu=p/q=p/m\). In the next two paragraphs, \(q\) will take the place
of \(m\) in order to facilitate contact with the literature. 

{\it \(m=q\) odd.---} 
For general \(p\) (including \(p=1\)), Berry phase calculations \cite{arovas1984,su1986}
determine the statistical angle associated to the exchange of clusters of \(p\) 
elementary quasiholes (or quasiparticles) to be \(\theta'=\pi p/q\). 
I translate this result obtained in first quantization into the exchange relations  
\begin{eqnarray*}
C_{\alpha_i}^pC_{\alpha_j}^p=e^{i\theta'}C_{\alpha_j}^pC_{\alpha_i}^p=
e^{\im \pi p/q}C_{\alpha_j}^pC_{\alpha_i}^p, \quad i<j,
\end{eqnarray*}
for clusters of \(p\) quasiholes. These relations are equivalent to the
additional constraint
\begin{eqnarray}\label{extrat}
e^{i\theta p^2}=e^{i\theta'}=e^{\im \pi p/q}
\end{eqnarray} 
on the fundamental statistical angle \(\theta\) of the fermion-root 
quasiparticles. As first noticed in Ref.\,\cite{su1986} in the first-quantized
framework of trial wave functions and Berry phase calculations, \(\theta\) 
is uniquely determined by Eq.\,\eqref{extrat} and the more fundamental 
constraint \(e^{i\theta q^2}=-1\). 
This is the procedure for computing the unique statistical angle that 
satisfies both exchange constraints: find any pair of integers \(n_1,n_2\) with 
\(n_1p+n_2q=1\); then,
\begin{eqnarray*}
\theta_{p/q}= \frac{\pi(n_1^2q+n_2^2p)}{q}\quad (\mbox{mod}\ 2\pi).
\end{eqnarray*}

There is an interesting special feature of the statistical angle
\(\theta=\pi/q\) associated to the Laughlin fractions \(\nu=1/q\):
In this case, the ``particle-hole" combinations 
\begin{eqnarray}\label{combs}
\Gamma_{\alpha_i}=C_{\alpha_i}+C_{\alpha_i}^{\dagger (2q-1)},\quad 
\Delta_{\alpha_i}=C_{\alpha_i} e^{\im \frac{\pi}{q} N_\alpha} +C_{\alpha_i}^{\dagger (2q-1)}
\end{eqnarray}
have well defined exchange statistics as well. In fact, as shown in 
Ref.\,\cite{cobanera2014}, the \(\Gamma_{\alpha_i},\Delta_{\alpha_i}\) 
satisfy the algebra of \(\mathds{Z}_{2q}\) parafermions \cite{alcaraz1981,jaffe2014,review}. 
For \(q=1\), that is, when there is no fractionalization, the unconventional 
Fock algebras of this paper reduce to the usual fermionic algebra, and 
Eq.\,\eqref{combs} yields Majorana fermions
\begin{eqnarray*}
\hspace{-1cm}
\Gamma_{\alpha_i}=C_{\alpha_i}+C_{\alpha_i}^\dagger\equiv a_{\alpha_i},\quad
\Delta_{\alpha_i}= 
C_{\alpha_i}(-1)^{C_{\alpha_i}^\dagger C_{\alpha_i}}+C_{\alpha_i}^\dagger
=-C_{\alpha_i}+C_{\alpha_i}^\dagger\equiv -\im b_{\alpha_i}.
\end{eqnarray*}
This connection between parafermions and Majorana fermions is important because 
it permits to extend the ideas of Ref.\,\cite{kitaev2001} in order to build a 
toy model of a  one-dimensional fractional topological superconductor. We will 
come back to this point in the next section. 

{\it \(m\) even.---}
The tight link between charge and anyon statistics is lost if \(m=2^rq\) 
is even \cite{su1986}. This is not to say that one cannot take an even root 
of a fermion. But there seems to be no physical grounds to prefer any one 
unconventional Fock algebra when \(m\) is even. We will make one more comment on 
this point below.
  
The last step in the complete determination of fermion-root quasiparticles
is to confront the question, {\it what is the exchange statistics between 
fractionalized and non-fractionalized fermions?} Suppose the system
of fractionalized electrons \(f_{\alpha_i}=C_{\alpha_i}^m\) coexists
with unfractionalized fermions \(c_{\beta_j}\). The creation and annihilation
operators \(c_{\beta_j}^\dagger,\ c_{\beta_j}\) create/annihilate 
ordinary, unfractionalized fermions in the single-particle states 
\(\beta_j\). Let us order the totality of the labels \(\alpha_i,\beta_j\)
so that \(\alpha\)-labels precede the \(\beta\)-labels.  Because of 
locality (we will see a concrete instance in the next section), fermions 
composite or elementary must anticommute,
\begin{eqnarray*}
\hspace{-1cm}
f_{\alpha_i}c_{\beta_j}+c_{\beta_j}f_{\alpha_i}=
C_{\alpha_i}^mc_{\beta_j}+c_{\alpha_j}C_{\beta_i}^m=0,\quad 
f_{\alpha_i}c_{\beta_j}^\dagger+c_{\alpha_j}^\dagger f_{\beta_i}=
C_{\alpha_i}^mc_{\beta_j}^\dagger+c_{\beta_j}^\dagger C_{\alpha_i}^m=0.
\end{eqnarray*}
This requirement introduces a new statistical angle \(\psi\) in the problem
such that
\begin{eqnarray*}
C_{\alpha_i}c_{\beta_j}=e^{i\psi}c_{\beta_j}C_{\alpha_i},
\quad C_{\alpha_i}c_{\beta_j}^\dagger=e^{-i\psi}c_{\beta_j}^\dagger C_{\alpha_i}
\end{eqnarray*}
and \(e^{im\psi}=-1\). I will not investigate \(\psi\) in detail. For 
odd \(m\), one can just require that the fermions and the fermion-root 
quasiparticles anticommute, 
\begin{eqnarray*}
C_{\alpha_i}c_{\beta_j}+c_{\beta_j}C_{\alpha_i}=0=
C_{\alpha_i}c_{\beta_j}^\dagger+c_{\beta_j}^\dagger C_{\alpha_i}
\qquad \quad \mbox{(allowed only for \(m\) odd)}.
\end{eqnarray*}
For even \(m\), there is no preferable answer. For example, for \(m=2\), 
the new exchange angle can be taken to be \(\psi=\pi/2\) so that  
\begin{eqnarray*}
C_{\alpha_i}c_{\beta_j}=\im c_{\beta_j}C_{\alpha_i},\quad
C_{\alpha_i}c_{\beta_j}^\dagger=-\im c_{\beta_j}^\dagger C_{\alpha_i}
\qquad \quad \mbox{(allowed for \(m=2\))}.
\end{eqnarray*}
In any case, it is easy to arrange for any of these condition to hold 
in a matrix realization of a mixed Fock algebra combining fractionalized
and unfractionalized fermions.

\section{Hybridization of Majorana and parafermion zero-energy edge modes by
charge-conserving tunneling}
\label{numerics}

\begin{figure}
\centering
\includegraphics[angle=0,width=0.9\columnwidth]{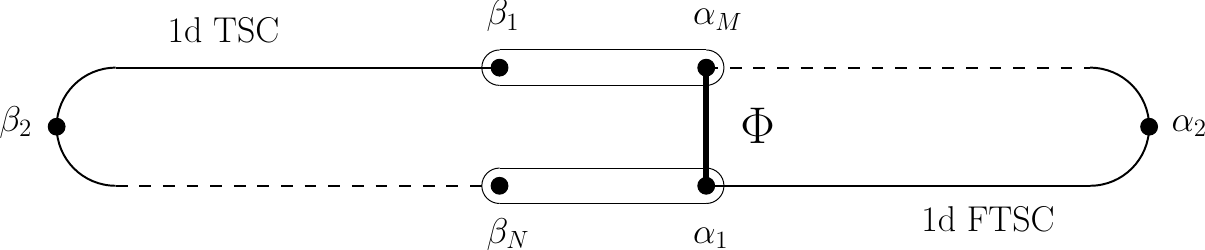}
\vspace{.2cm}

\includegraphics[angle=0,width=.45\columnwidth]{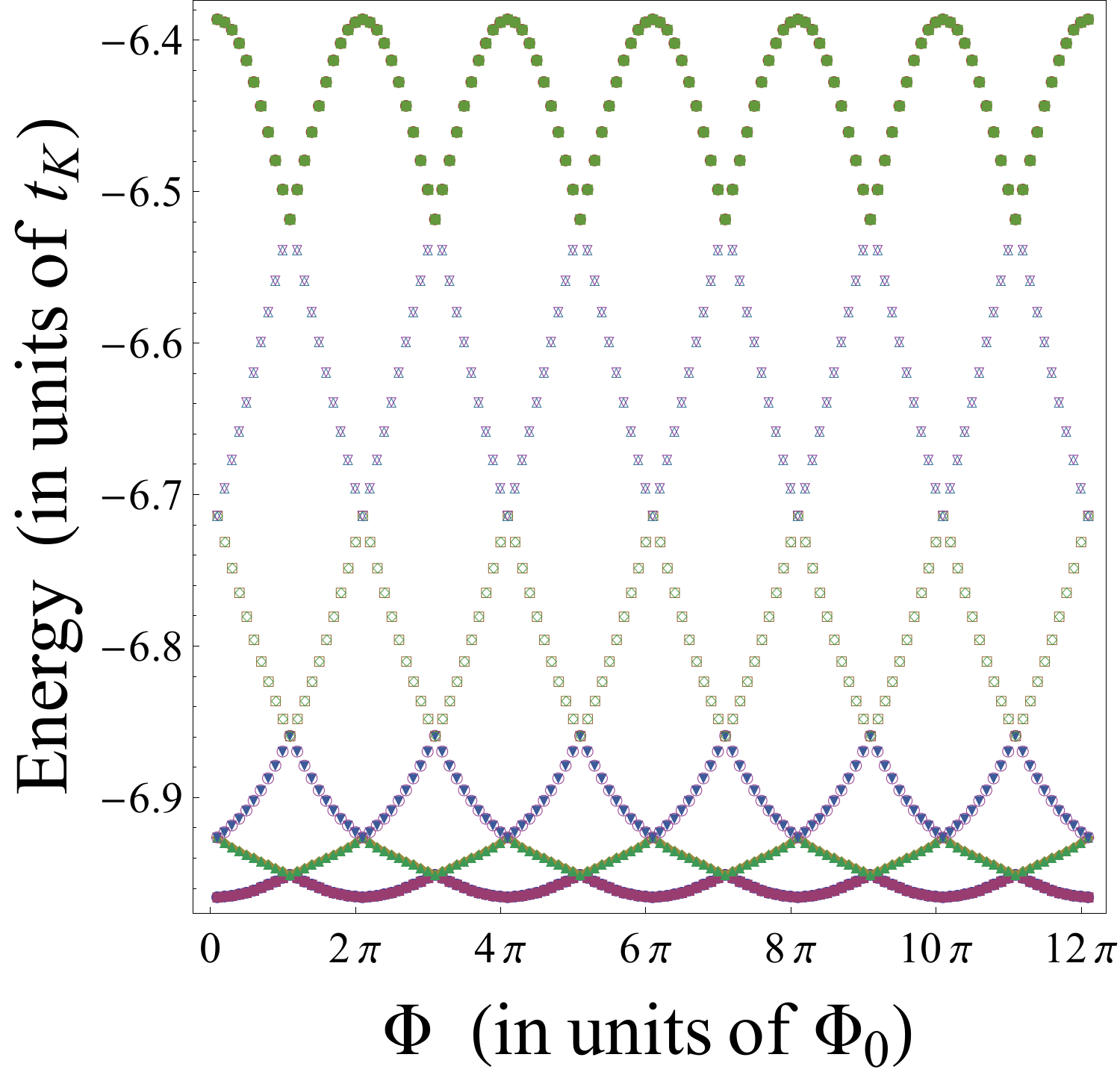}
\includegraphics[angle=0,width=.45\columnwidth]{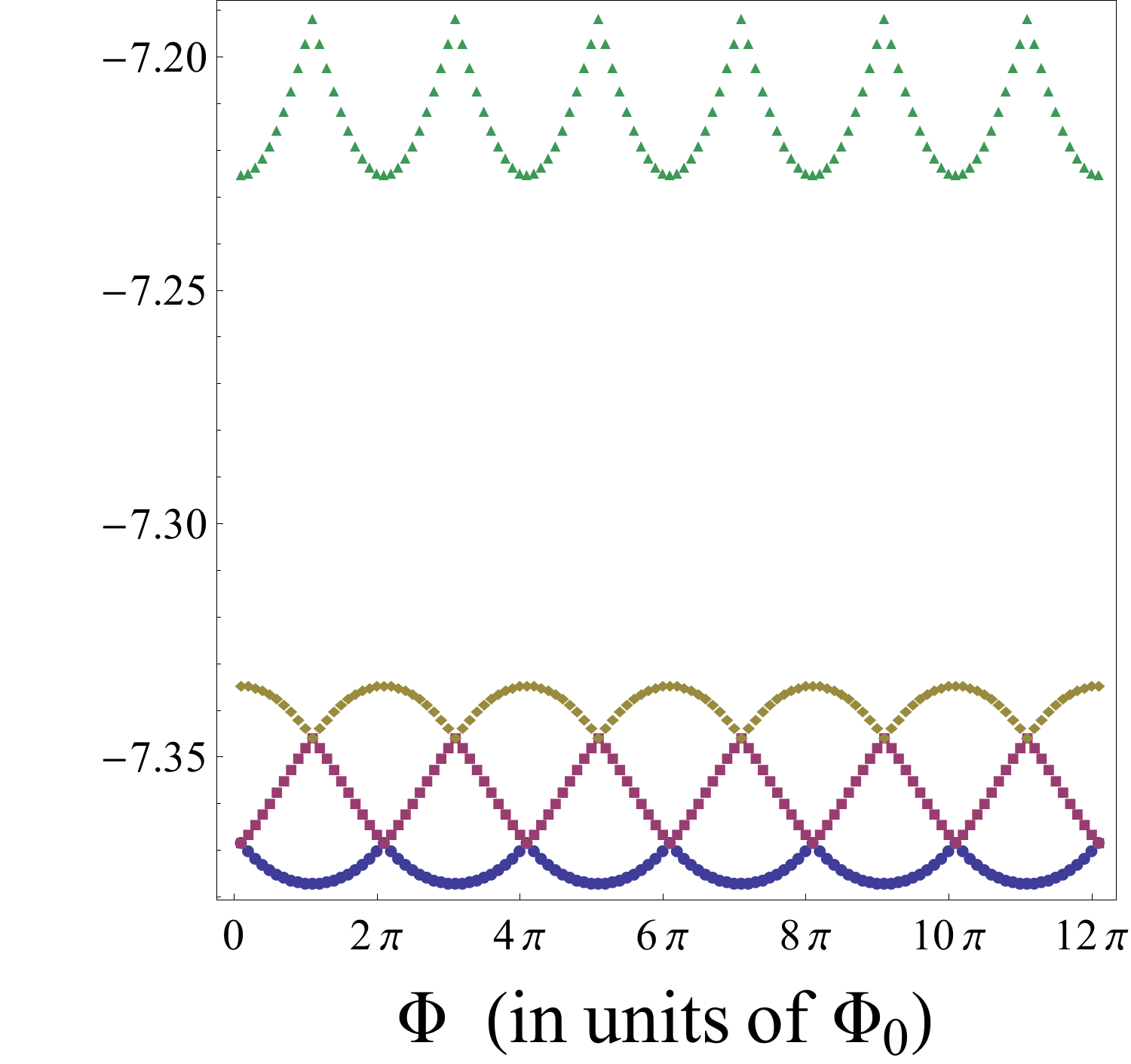}
\caption{
Hybridization of Majorana and parafermion zero-energy edge modes. 
{\it Top panel:} Schematic representation of the model Hamiltonian: 
The endpoints of a Majorana chain (D=1 topological superconductor (TSC)) 
and the endpoints of of a parafermion chain (D=1 fractional topological 
superconductor (FTSC)) are connected by fractionalizing, charge-conserving 
tunneling. A fractionalizing link closes the resulting 
hybrid wire into a ring junction, threaded by a flux \(\Phi\) 
measured in units of \(\Phi_0=hc/2e\), see Eq.\,\eqref{that}. 
In the following, \(m=3\), \(t_K=t_P=t=1\), and \(M=N=3\); see
main text. 
{\it Bottom left panel:} 
If the chains are decoupled, exact diagonalization of the junction 
Hamiltonian Eq.\,\eqref{hj} shows that the six lowest energy levels 
are doubly degenerate and do not cross higher levels. The period of 
the equilibrium supercurrent \(I_{sc}\propto dE/d\Phi\) is \(12\pi\), 
consistent with \(\mathds{Z}_6\) parafermion edge modes. 
{\it Bottom right panel:} 
If the chains are coupled by tunneling, exact diagonalization shows 
that the three lowest energy levels of the junctions are non-degenerate 
and do not cross higher levels. The period of the supercurrent is 
\(6\pi\), consistent \(\mathds{Z}_3\) parafermion edge modes.}
\label{abridge}
\end{figure}

Now the specification of the fermion-root quasiparticles is complete 
and so a concrete application is due. In this section I will consider
a system of \(N\) fractionalized spinless fermions \(f_i=C_i^m\)
hopping on a chain with sites \(i=1,\dots,N\), and \(M\) unfractionalized
fermions \(c_j\) hopping on another chain with sites \(j=1,\dots,M\),
see Fig.\,\ref{abridge}. The Hamiltonian for the unfractionalized 
fermions is 
\begin{eqnarray*}
H_K=-t_K\sum_{j=1}^{M-1}(-\im b_j)a_{i+1}= 
-t_K\sum_{j=1}^{M-1} (-c_i+c_i^\dagger)(c_{i+1}+c_{i+1}^\dagger).
\end{eqnarray*}
This is the Hamiltonian for the Majorana chain of Kitaev \cite{kitaev2001}
deep in its topologically nontrivial regime, with perfectly decoupled Majorana 
zero-energy edge modes \(a_1,b_M\). No topological number has been computed 
for the parafermion generalization  
\begin{eqnarray*}
\hspace{-1cm}
H_{\sf P}=-t_P\sum_{i=1}^{N-1}(\Delta_i\Gamma_{i+1}^\dagger+H.c.)=
-t_P\sum_{i=1}^{N-1}(C_{i} e^{\im \frac{\pi}{q} N_\alpha} +C_{i}^{\dagger (2q-1)})
(C_{i+1}+C_{i+1}^{\dagger (2q-1)}).
\end{eqnarray*}
Nonetheless, \(H_{\sf P}\) hosts zero-energy edge modes \(\Gamma_1,\Delta_N\) 
that are protected simply because no relevant perturbation exists that is 
local in terms of quasiparticle operators \cite{foot5} (see Ref.\,\cite{fendley2012}
for a different point of view). The full Hilbert space for both chains combined 
is of dimension \(2^M(2m)^N\).

Let us put the edges of the two chains in contact with tunneling Hamiltonians 
\begin{eqnarray}\label{frach}
H_{1}=-\gamma(c_M^\dagger C_{1}^m-c_{M}C_{1}^{\dagger m}),\quad
H_{2}=-\gamma(c_1^\dagger C_{N}^m-c_{1}C_{N}^{\dagger m}).
\end{eqnarray}   
They commute with each other and they commute with the 
total-charge operator 
\begin{eqnarray*}
Q_{\rm total}=e^*\sum_{i=1}^N \sum_{l=1}^{2m-1}C_i^{\dagger\, l}C_i^l
+e\sum_{j=1}^M c^\dagger_{j}c_{j}=\frac{e}{m}\sum_{i=1}^N N_i+e\sum_{j=1}^Mn_j.
\end{eqnarray*}
If the tunneling amplitude \(\gamma\) for charge breakup processes does 
not vanish, then the edge modes of the two chains will hybridize. To 
investigate the edge-modes that may emerge as a consequence, let us 
introduce a quasiparticle hopping term 
\begin{eqnarray}\label{that}
H(\Phi)=-t(C_MC_1^{\dagger} e^{-\im \Phi/2m}+C_1C_M^\dagger e^{\im \Phi/2m})
\end{eqnarray}
that closes the system into a short ring junction. Like the tunneling
Hamiltonians, \(H(\Phi)\) commutes with the total charge operator. The 
phase \(\Phi\), measured in units of the superconducting flux quantum 
\(\Phi_0=hc/2|e|=hc/2m|e^*|\), induces an
equilibrium Josephson current through the ring.
The total Hamiltonian of the ring junction is 
\begin{eqnarray}\label{hj}
H_{\rm junction}=H_K+H_P+H_{1}+H_2+H(\Phi).
\end{eqnarray}

For \(m=3\), exact diagonalization of the junction Hamiltonian 
shows that the the period of the equilibrium Josephson current in this 
is \(6\pi\), see Fig.\,\ref{abridge}. 
Hence, the hybridization of the Majorana and \(\mathds{Z}_6\) parafermion 
edge modes generates two \(\mathds{Z}_3\) parafermion edge modes. But, 
unfortunately \cite{mong2014}, the host system is not critical. 
Ref.\,\cite{foot5} describes a way of realizing a \(\mathds{Z}_3\) parafermion 
chain on a critical line.

\section{The \(m\)th root of a fermion carrying a non-Abelian charge}

The principal attribute of fermion-root quasiparticles is that they carry
a fraction of the electric charge of the underlying fermions. But what if 
the fermions carry also non-Abelian charges? Non-Abelian charges are quantized 
and tied to an intrinsic scale set by the commutation relations of the 
non-Abelian charges. They cannot be fractionalized or renormalized in 
the same way Abelian charges can be. This observation is perplexing because 
fermion-root quasiparticles carry the same single-particle labels that 
characterize the underlying fermions. It must be that the non-Abelian 
quantum numbers carried by the labels \(\alpha\) are an attribute of the 
composite fermion \(C_\alpha^m\), and not of the quasiparticle \(C_\alpha\). 
So what are the non-Abelian quantum numbers carried by the fermion-root 
quasiparticles? Next I will provide a natural, but possibly non-unique in
a physical sense, answer to this question. For concreteness I will focus 
on systems of spinful electrons. The generalization to other internal, 
compact non-Abelian symmetries is straightforward. 

For spinful electrons, the single-particle labels are pairs \(\r,\sigma\) 
consisting of a lattice site \(\r\) and 
the spin component \(\sigma=\uparrow,\downarrow\) along the quantization axis.
Now imagine a spin-conserving system split into a normal (non-fractionalizing)
region with Hamiltonian \(H_N\) (lattice sites \(\x\)), a fractionalizing 
region \(H_F\) (lattice sites \(\y\)), and an interface \(H_I\). The interface 
Hamiltonian contains only terms of the form of Eq.\,\eqref{frach}. If I at first
I take \(H_F=0\), then the total Hamiltonian for the system is \(H_N+H_I\), 
and the total spin operator is   
\begin{eqnarray*}
J^a=\sum_{\x} s_{\x}^a+\sum_{\y} S_{\y}^a,\quad a=x,y,z.
\end{eqnarray*} 
For the normal region, the local spin is given as usual
by \(s^-_{\x}=f^\dagger_{\x,\downarrow}f_{\x,\uparrow},\ 
s^z_{\x}=(n_{\x,\uparrow}-n_{\x,\downarrow})/2\). For the
composite fermions at the interface, the local spin is 
\begin{eqnarray}\label{sf}
S^-_{\y}=C_{\y,\downarrow}^{\dagger m}C_{\y,\uparrow}^m,\ \
S^z_{\y}=\frac{1}{2}(C_{\y,\uparrow}^{\dagger m}C_{\y,\uparrow}^m
-C_{\y,\downarrow}^{\dagger m}C_{\y,\downarrow}^m).
\end{eqnarray}
Now rather than asking what is the correct \(J^a\) for a non-trivial 
\(H_F\), I propose that \(J^a\) is always correct. Then the question is,
can one build model Hamiltonians \(H_F\) that conserve spin and 
display deconfined fractional charge? The answer is in the affirmative, 
because the total charge per site 
\(Q_{\y}=e^*(N_{\y,\uparrow}+N_{\y,\downarrow})\) 
commutes with the the spin operators of Eq.\,\eqref{sf},
\begin{eqnarray*}
[Q_{\y},S^-_{\y'}]=0=[Q_{\y},S^z_{\y'}].
\end{eqnarray*}
Hence, one can construct an alternative set of quasiparticles that carry 
fractional charge and definite values of spin.  
These are the elementary excitations of the fractionalizing non-Abelian 
medium, and the fundamental objects for building model Hamiltonians. 

Let me denote by \(L_k=(l_1,l_2,l_3,l_4)\) the set of \(l_i=0,\dots,2m-1\)
such that \(l_1-l_2+l_3-l_4=k\), with \(k\) an integer \(|k|\leq 2(2m-1)\).
Then the combinations \(C_{\y}^{L_k}=C_{\y,\uparrow}^{\dagger l_1}C_{\y,\uparrow}^{l_2},
C_{\y,\downarrow}^{\dagger l_3}C_{\y,\downarrow}^{l_4}\) change the
local amount of charge by \(ke^*\) units,
\begin{eqnarray*}
[Q_{\y'},C^{L_k}_\y]=ke^*C^{L_k}_\y\delta_{\y,\y'}. 
\end{eqnarray*}
The excitations created by these combinations are not eigenstates of spin in 
general. But because spin commutes with the local charge, a rotation in spin 
space does not mix different \(k\). Hence one obtains a linear transformation 
\begin{eqnarray*}
\, [S_{\y}^a,C_{\y}^L]&=&\sum_{L_k'} (j^a)_{L_k'}^{L_k}C_{\y}^{L_k'}.
\end{eqnarray*}
From this point on one follows the standard theory of tensor operators. 
The angular momentum matrix \(j^z\) can be diagonalized by introducing 
linear combinations of the \(C_\r^{L_k}\). These new quasiparticles 
\(A_\y^K\), \(K=(k,s,m_s)\), carry charge \(ke^*\), and spin \(s(s+1)\) with 
projection \(m_s\) onto the quantization axis. The \(A\) quasiparticles
have well defined exchange statistics since
\begin{eqnarray*}
A_{\y_1}^{K_1} A_{\y_2}^{K_2}= e^{i\theta k_1k_2}A_{\y_2}^{K_2} A_{\y_1}^{K_1}.
\end{eqnarray*}
A spin-conserving hopping term reads 
\(A_\y^KA_{\y'}^{K \dagger}+A_{\y'}^{K}A_\y^{K \dagger}\).

The change of basis from the \(C\) to \(A\) quasiparticles is 
computationally inexpensive because it is local: all one has to do 
is to compute the change of basis for a system with only one site. 
Then, the dimension of the relevant Fock space is \(4m^2\), and the 
matrices \(j^a\) are of small dimension. Moreover, they are easy to 
compute because the \(C^{L_k}\) are orthogonal with respect 
to the trace inner product. I will not show the outcome of such a 
systematic calculation here. I will present one spin \(1/2\) multiplet 
of fractionally-charged quasiholes to illustrate some counterintuitive 
features. The \(A\) quasihole
\begin{eqnarray*}
A_{\y}^{\downarrow}= C^{\dagger 2}_{\y,\uparrow}
C_{\y,\uparrow}^3-C^{\dagger 3}_{\y,\uparrow}C_{\y,\uparrow}^4
\end{eqnarray*}
of fractional charge \(-e^*\) features no ``spin" down \(C\) quasiparticles.
Nonetheless, it lowers the angular momentum along the quantization axis,
since \([S^z_\y,A_\y^\downarrow]=-(1/2) A_\y^\downarrow\). Also, it is 
a member of a spin \(1/2\) multiplet because \([S^-_\y,A_\y^\downarrow]=0\). 
Hence, one can compute the other member of the multiple as 
\(A_\y^\uparrow=[S_\y^+,A_\y^\downarrow]\), which yields
\begin{eqnarray*}
A_{\y}^{\uparrow}=-C^{\dagger 2}_{\r,\uparrow}C_{\r,\downarrow}^3+
C^{\dagger 3}_{\r,\uparrow}C_{\r,\uparrow}C_{\r,\downarrow}^3-
C^{\dagger 5}_{\r,\uparrow}C_{\r,\uparrow}^3C_{\r,\downarrow}^3
\end{eqnarray*}
and check explicitly that \([S^z_\y, A_{\y}^{\uparrow}]=(1/2)A_{\y}^{\uparrow}\) and 
\([S_\y^+,A_\y^\uparrow]=0\). The adjoint spin \(1/2\) multiplet 
\(A_\y^{\downarrow \dagger}, A_\y^{\uparrow \dagger}\) creates quasiparticles 
of charge \(e^*\).

\section{Summary and outlook}

The procedure of taking the \(m\)th root of a fermion yields \(m^2\) 
unconventional Fock algebras. The associated quasiparticles satisfy 
\(2m\)-exclusion statistics, that is, Eq.\,\eqref{exc}, and the normal 
ordering rule Eq.\,\eqref{no}. These two properties suffice to enforce 
the key result that the number operator defined in Eq.\,\eqref{counting} 
generates \(U(1)\) rotations of the fermion-root quasiparticles, see 
Eq.\,\eqref{u1}. The only feature that distinguishes the various types 
of fermion-root quasiparticles is the statistical angle 
\(\theta_\ell=\pi(2\ell+1)/m^2,\ \ell=1,\dots,m^2-1\). If the system 
under investigation features both fractionalized and unfractionalized 
fermions, then, in order to respect locality, one must demand that the 
``composite" fermions obtained from filling a single-particle state with 
\(m\) fermion-root quasiparticles anticommute with the unfractionalized 
fermions. This condition introduces a new statistical angle \(\psi\) in 
the problem for characterizing the result of exchanging an unfractionalized
fermion with a fermion-root quasiparticle. There are simple concrete matrix 
realizations of these unconventional Fock algebras. Finally, if the fractionalized 
fermions carry non-Abelian charges, then then the fermion root 
quasiparticles do not carry well defined values of these charges. But
one can make a rotation in charge space in order to obtain quasiparticles
that carry fractional charge and well defined values of the non-Abelian
charges. 

I conclude with some ideas for future research. One possibility would be
to model other mesoscopic(-inspired) systems like the hybrid chain of 
this paper. The study of mesoscopic devices that include fractionalizing 
components is an active field in connection to topological quantum 
computation \cite{lindner2012,burrello2013,pfsbib,mong2014}.
Using the quasiparticles of this paper it would be straightforward,
thanks to advances in computational many-body physics like tensor networks,
to study various forms of fractionalized Andreev bound states numerically \cite{bena2012}. 
Another possibility would be to study the shot noise spectrum of pinched 
fractional edge channels, at moderate values of the effective inter-edge 
transmission where the Luttinger liquid approach becomes intractable (see 
Ref.\,\cite{heiblum2009}, and references therein).

\section*{Acknowledgements}

I would like to gratefully acknowledge discussions with
Denis Chevallier, Wellington Galleas, Bernard van Heck, Gerardo Ortiz, 
Dirk Schuricht, and Cristiane Morais Smith. This work is part of the DITP consortium, 
a program of the Netherlands Organisation for Scientific Research (NWO) 
that is funded by the Dutch Ministry of Education, Culture and 
Science (OCW).

\end{document}